# Prediction of ultrasonic cavitation with a dimensionless number, towards higher reproducibility.

Name: Gonzalo     Family name: Garcia-Atance Fatjo

Jost Institute for Tribotechnology, University of Central Lancashire, Preston PR1 2HE, UK

Email: ggarcia-atancefatjo@uclan.ac.uk
https://orcid.org/0000-0002-3914-7160

**Abstract**

Ultrasonic cavitation is the formation of vapour cavities within a liquid due to the action of an ultrasound source. It is widely used for homogenization, dispersion, deagglomeration, erosion, cleaning, milling, emulsification, extraction, disintegration and sonochemistry. On the other hand, the so-called cavitation number is used to assess the likelihood of cavitation in fluid flows within a conduit or around a hydrofoil but it is not valid in ultrasonic cavitation since there is no fluid flow. A recently formulated number predicts the cavitation in case of sudden accelerations. The tip surface of an ultrasonic probe is subjected to a continuous repetition of alternating accelerations at high frequency. Therefore, the use of the recently formulated number in ultrasonic cavitation is explored here. Simulations of the ultrasonic probe in water just at the condition of cavitation onset have been performed for a combination of probe diameters from 0.2 to 100 mm and frequencies 20, 30, 40, 100 and 1000 KHz. The recently formulated number is applied to these combinations and it is found that can be used to predict ultrasonic cavitation. Consequently, the dimensionless number can be used to decide the conditions to avoid or generate cavitation when a fluid is sonicated and to increase reproducibility in such conditions.

1. Introduction

Vaporous cavitation in a fluid is the change from liquid phase to vapour phase due to a drop in the local pressure, as opposed to boiling where the change of phase is due to a rise in temperature. As the change of pressure propagates much faster than the change of temperature, the cavities can be generated and collapsed rapidly producing turbulence, sound, pressure shockwaves, local rise of temperature and surface erosion. Thoma and Leraux introduced in 1923-1925 a dimensionless number called cavitation number (1) that is used in fluid flows, for example, to either prevent cavitation in a hydrofoil or in a conduit, or to produce cavitation in laboratory tests. This number cannot be applied to ultrasonic cavitation.

Ultrasonic cavitation is widely used for homogenization, dispersion, deagglomeration, erosion, cleaning, milling, emulsification, extraction, disintegration and sonochemistry. Different processes require different intensities of cavitation. The way to control the intensity is typically with the power output and frequency of the ultrasonic probe. However, the power output has limitations to indicate the existence of cavitation. If the pressures created by the sound wave are not enough to reduce the local pressure below the vapour pressure, the sound wave could be transmitted to the liquid without generation of cavitation.

There is a wide range of sonication intensities that are commonly used. This range goes from high intensity cavitation to ultrasound without cavitation. The absence of cavitation during sonication was proven useful in some chemical reactions (2) and in increasing the temperature of the nucleation of ice in water (3). Low acoustic intensity can enhance bacteriological processes (4, 5). This is especially important in sonobioreactors, where excessive cavitation can damage the cell culture, although ultrasound produces beneficial effects (6). On the other hand, there are applications where ultrasound is used but cavitation is not desirable such as medicine or ultrasound diagnosis (7)

Sonochemistry has been proven to have in many occasions low reproducibility due to geometrical differences of the beaker or reactor, and power and frequency dependence on the probe itself (8). On the other hand, in biology, the failure to reproduce the results from another investigator during sonication of cells in vitro is common due to the variability in the occurrence of cavitation (9). Mason and Peters listed a series of parameter that should be reported in order to increase the poor reproducibility commonly found (10). Among those parameters are frequency, intensity and shape. Researchers from the National Physical Laboratory reported the creation of a reference cavitating vessel to increase reproducibility in industrial and research applications (11, 12). Consequently, a dimensionless number that quantifies the existence of cavitation could be a complementary parameter that increases accuracy in the replication of conditions in research and industrial applications, especially when the conditions are close to the cavitation threshold.

Therefore, a dimensionless number that predicts the existence or absence of cavitation and the proximity to the threshold between the two conditions is a contribution with immediate application that can increase reproducibility. This article provides this dimensionless number applied to ultrasonic cavitation.

## 2. Dimensionless number.

In 2016 a new dimensionless number to predict cavitation in accelerated systems was reported (13). It was applied to the "tube arrest" method and to the downstream flow of a closing valve. The new dimensionless number was formulated as

$$G_A = \frac{p - p_v}{\rho a L} \tag{1}$$

where $G_A$ is the new dimensionless number, $p$ is the pressure of the liquid before or after being subjected to acceleration, $p_v$ is the vapour pressure of the liquid, $\rho$ is the density of the liquid, $L$ is the effective length of the liquid or column subjected to acceleration measured parallel to the acceleration vector, and $a$ is the acceleration.

In order to apply this number to an ultrasonic probe, the acceleration $a$ is calculated as a function of the frequency and the amplitude. The length $L$ of the liquid subjected to acceleration has to be estimated.

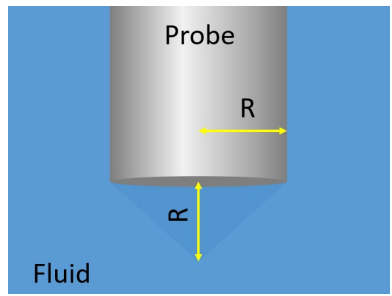

Figure 1 **Estimation of the length of water under direct influence of the acceleration of the tip surface.**

Figure 1 shows an estimation of the length of water under direct influence of the movement of the tip. The length $L$ is estimated to be the radius of the probe. This kind of estimations are typical in the utilisation of dimensionless numbers. The justification for this estimation is that it is a representative size of the system with physical meaning, easy to remember, and resembles the cavitation field normally seen in these systems.

The length $L$ is, as stated in (13), the shortest of the geometrical length, in this case the radius $R$, and the result of $c\Delta t$ where $c$ is the speed of sound and $\Delta t$ is the time that the acceleration lasts. $\Delta t$ is taken for simplicity half the period, since half of the period the acceleration is producing a tension pulse and the other half is producing a compression pulse. This approach is justified later in the results.

In order to calculate the acceleration $a$ of the tip surface a sine movement of the tip is assumed with frequency $f$ and amplitude $d$. The maximum acceleration is given by:

$$a = d\,(2\,\pi\,f)^2 \tag{2}$$

The recently presented dimensionless number $G_A$ will indicate the likelihood of cavitation within the liquid. In general if $G_A \ll 1$, the liquid will cavitate. This is set in a similar manner as the traditional cavitation number that will indicate cavitation when $N_{CA} \ll 1$. However, the traditional one is meaningless in ultrasonic cavitation as shown below. It is given by the equation:

$$N_{CA} = \frac{p - p_v}{1/2\, \rho v^2} \qquad (3)$$

where $N_{CA}$ is the traditional cavitation number, $p$ is the pressure of the undisturbed liquid, $p_v$ is the vapour pressure of the liquid, $\rho$ is the density of the liquid and $v$ is the velocity of the liquid, for the ultrasonic case, we take the value of the speed of the ultrasonic probe tip since the rest of the fluid is still.

Let's assume a condition where it is known that cavitation happens, for example an ultrasonic probe of 15.9mm diameter oscillating at 20 KHz with an amplitude of 5 μm in water at room conditions. For this situation $N_{CA} = 495$ that is much greater than 1 and suggest that there is no cavitation demonstrating that $N_{CA}$ is meaningless for ultrasonic cavitation as it was already known. Conversely $G_A = 0.156$ that is much smaller than 1 and indicates that there is cavitation as it is the case. Other comparisons between $N_{CA}$ and $G_A$ can be found in (13).

3. **Simulation set up.**

Many commercial ultrasonic probes produce cavitation in all working conditions, even when there are means of reducing the power. Therefore, in order to investigate the validity of $G_A$, a simulation of the ultrasonic probe for multiple diameters, amplitudes and frequencies is chosen.

The simulation is made with FEM by coupling compressible laminar flow with acoustics using a moving mesh. The geometry is a cylindrical probe placed in the centre of a cylindrical beaker. The beaker has a flat bottom with an acoustic impedance of $12.3 \cdot 10^6$ Pa·s/m. The speed of sound in water is taken as 1481 m/s. The tensile strength of water is assumed to be null, thus there are enough nuclei within the liquid that create the conditions for cavitation when parts of the liquid go into negative absolute pressure. This reasoning is also embedded in the formulation of the traditional cavitation number.

The simulation time was long enough to allow the pressure wave to be reflected on the wall of the beaker and come back to the probe tip. At 20Khz this is achieved easily since the period is almost as long as the time of flight of the sound wave. In Figure 2 the wave generated by an ultrasonic probe at 20 Khz is shown during the first cycle just before arriving to the bottom of the beaker. The wave is almost spherical. It losses intensity the longer it travels. Therefore, the influence of the reflected wave on the minimum pressure is limited. Conversely, at higher frequencies, for example at 1 MHz, the wavelength is much smaller than the diameter of the probe. Consequently, the wave is almost longitudinal, its intensity does not decrease so much and the influence of the reflection at the bottom of the beaker is going to change the pressure field considerably, see Figure 3. The results presented here include the reflection of the pressure wave in the beaker unless otherwise indicated.

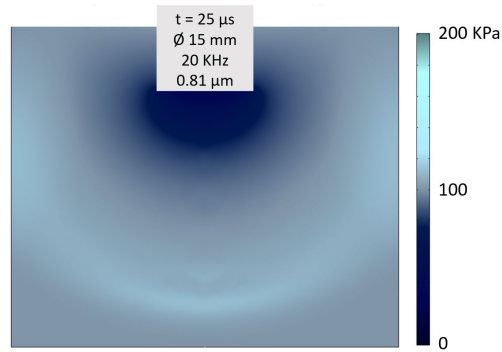

Figure 2 **Pressure field in a beaker at 25µs, during the first cycle, subjected to sonication with a 15 mm diameter probe, at 20KHz, with amplitude 0.81 µm. The compressive wave is about to reach the bottom of the beaker and being reflected.**

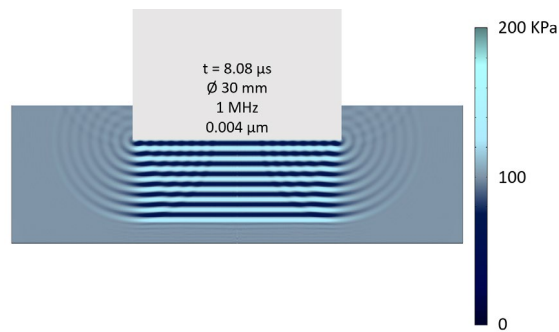

Figure 3 **Pressure field in a beaker at 8.08µs, approximately at the 8$^{th}$ cycle, subjected to sonication with a 30 mm diameter probe, at 1 MHz, with amplitude 0.004 µm. The pressure waves are about to reach the bottom of the beaker and being reflected.**

4. Results.

The results are presented in charts amplitude versus diameter. See Figures 4-8. For each probe diameter, there is an amplitude limit above which cavitation exists. Conversely for smaller amplitudes cavitation does not occur. It is easy to conclude that the region of the chart with high amplitude corresponds to the region where cavitation happens while the region with low amplitudes cavitation does not happen. The shape of the transition between this two regions and the location of such transition is calculated with the finite element method and represented by square dots in the chart. At the same time the contourlines for $G_A$ values of 0.5, 1 and 2 are included in these charts. $G_A$ is calculated using the radius of the probe for the left part of the charts. For the right end, where diameters of the probe are big, and the radius of the probe is longer than half the wavelength, half the wavelength is used. Half the wavelength is calculated as $c\Delta t$ where $c$ is the speed of sound and $\Delta t$ is half the period. This explains why $G_A$ is not dependent on the diameter of the probe when the dimeter is large. It becomes a horizontal line on the right part of the chart.

The selection of frequencies used for this study are 20 KHz (Figure 4), 30 KHz (Figure 5), 40 KHz (Figure 6), 100 KHz (Figure 7) and 1 MHz (Figure 8). The higher the frequency is, the shorter the wavelength is, and therefore, the longer the horizontal part of the countourline for $G_A$ is.

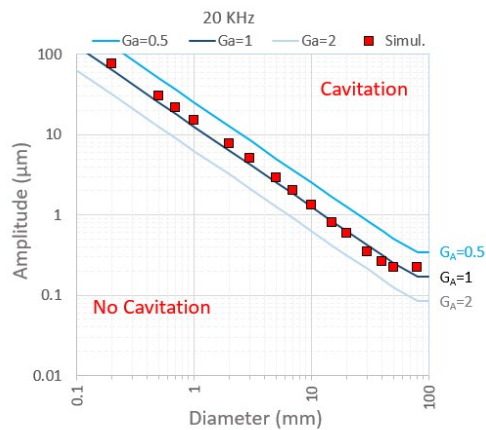

Figure 4 **Amplitude needed to cavitate for a given probe diameter at 20 KHz in water at room conditions. Contourlines for $G_A$ values of 0.5, 1 and 2 compare with the onset of cavitation according to simulations represented as squares.**

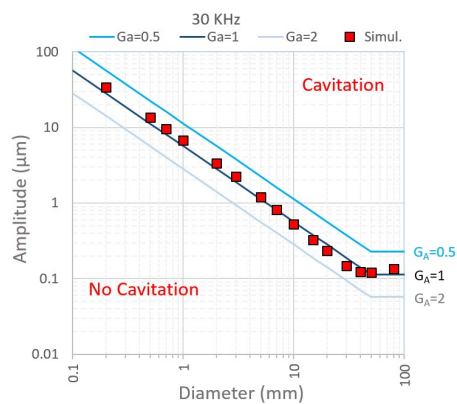

Figure 5 **Amplitude needed to cavitate for a given probe diameter at 30 KHz in water at room conditions. Contourlines for $G_A$ values of 0.5, 1 and 2 compare with the onset of cavitation according to simulations represented as squares.**

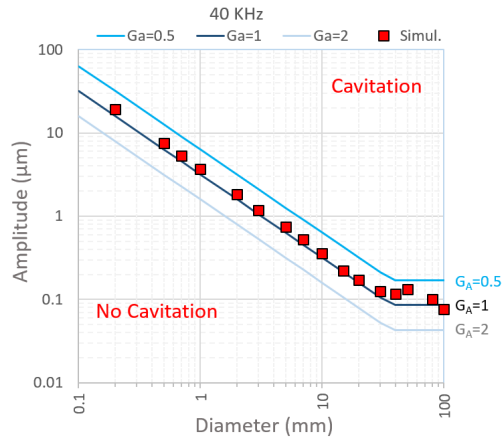

Figure 6 **Amplitude needed to cavitate for a given probe diameter at 40 KHz in water at room conditions. Contourlines for $G_A$ values of 0.5, 1 and 2 compare with the onset of cavitation according to simulations represented as squares.**

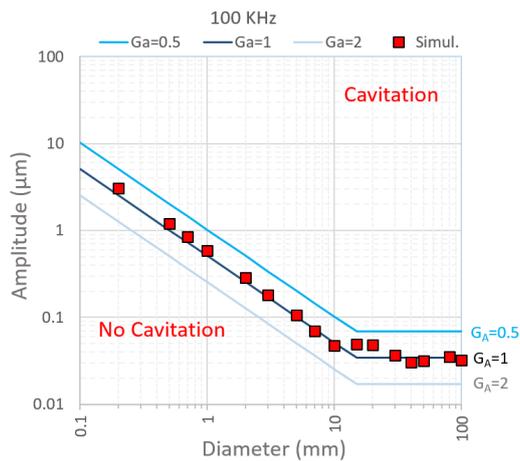

Figure 7 **Amplitude needed to cavitate for a given probe diameter at 100 KHz in water at room conditions. Contourlines for $G_A$ values of 0.5, 1 and 2 compare with the onset of cavitation according to simulations represented as squares.**

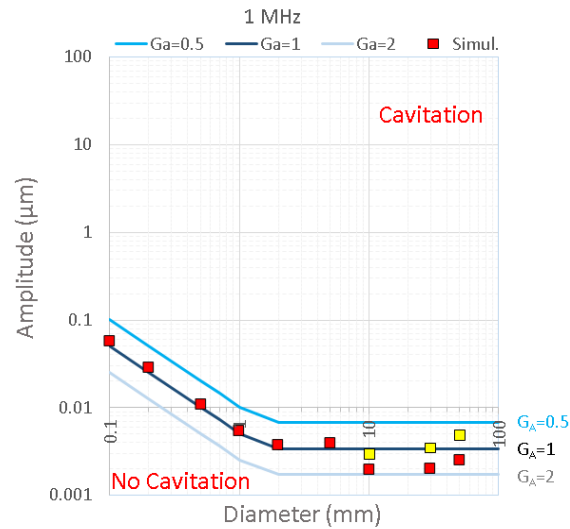

Figure 8 **Amplitude needed to cavitate for a given probe diameter at 1 MHz in water at room conditions. Contourlines for $G_A$ values of 0.5, 1 and 2 compare with the onset of cavitation according to simulations represented as squares. Yellow squares indicate without reflection of wave pressures.**

5. Discussion.

Figure 4 shows the results for 20 KHz. The red squares represent the condition of amplitude and probe diameter of the cavitation onset. At higher amplitudes the liquid would cavitate, at lower amplitudes the liquid will not cavitate. The location of these squares coincide remarkably well with the contourline for $G_A = 1$ . At high diameters close to 100mm the contourline for $G_A = 1$ present a deflection and becomes horizontal. This behaviour is also shown by the simulation with diameter 80 mm and repeated at higher frequencies in the other figures.

The results for 30 KHz, 40KHz and 100 KHz are very similar with the corresponding shift for the values (Figure 5, 6 and 7) due to the change of frequency. The simulations show that the threshold for cavitation is very close to $G_A = 1$ and always well within 0.5 and 2.

For 1 MHz, (Figure 8), the fitting of the simulations for small diameters to the contourline $G_A = 1$ is very good. However, for bigger diameters, where half of the wavelength is shorter than the radius of the probe, there is a departure of the contourline $G_A = 1$ towards greater values of $G_A$ . It means that for bigger diameters and high frequencies it is easier to get regions of the fluid with lower pressures, due to the reflection of the very directional soundwaves in the beaker walls. Consequently, a series of simulations without allowing time for the soundwaves to be reflected on the walls have been performed and are represented in Figure 8 as yellow squares. This yellow squares are closer to the contourline $G_A = 1$ .

In the deflection, or change from tilted to horizontal, of the contourline shown in the charts there is a small departure from $G_A = 1$ of the simulations. This is easy to observe in Figure 6 (40KHz) and Figure 7 (100 KHz) where the only two squares that are not actually touching $G_A = 1$ are in this region, above the line $G_A = 1$. The meaning of this is that at these diameters it is slightly more difficult to get cavitation and the amplitude has to be slightly bigger. This behaviour is attributed to the propagation of waves from the border of the probe. This waves are clearly visible in Figure 3 although for a case with higher frequency. The centre of these toroidal waves are the two lower corners, left and right, of the probe. Since Figure 3 depicts a cross section of the probe, these waves have a circular shape with centre on the left and right lower corners of the probe. These waves seem to couple with the main waves and reduce the pressure oscillations for these diameters where the wavelength is similar to the diameter size.

In general, the contourline $G_A = 1$ defines with high degree of accuracy the threshold for cavitation. Given a frequency, the threshold is dependent on the amplitude for small diameters, however the threshold is not dependent on the amplitude for relatively big diameters. A further investigation of this phenomena using a simple calculation of the radiated power in each case is developed.

The power radiated per unit of area is a parameter often used in ultrasonic cavitation to quantify the intensity of cavitation. Using a simple model to calculate power

$$P = F \cdot v \tag{4}$$

where $P$ is the power, $F$ is the force and $v$ is the velocity. Applying this model to a small area in front of the probe tip we can have

$$\frac{P}{A} = p_g \cdot v \tag{5}$$

where first term is the power radiated per unit of area, $p_g$ is the gauge pressure and $v$ is the velocity of the probe tip. As the gauge pressure is constant and equal to $p - p_v$ for the condition of cavitation onset, it is concluded that the power radiated per unit of area is proportional to the tip velocity for the condition of cavitation onset.

The maximum tip velocity is given by the equation

$$v = d(2\pi f) \tag{6}$$

therefore the tip velocity is proportional to the amplitude $d$ for a given frequency $f$.

The consequence of equations 5 and 6 is that the power radiated per unit of area for the condition of cavitation onset is proportional to the amplitude. As it is shown in figures 4 to 8, the cavitation onset for a given frequency could depend on the amplitude if the diameter of the probe is relatively small. For very high frequencies and diameters the cavitation onset happens at constant values of amplitude, therefore in that condition and only in that conditions the power radiated per unit of area is a good parameter to control the cavitation occurrence. However, in many normal conditions with frequencies of 20 to 30 KHz and diameters smaller than 80 and 50mm respectively, cavitation onset would happen at different values of amplitude and therefore with different powers per unit of area. For example at 20 KHz and with diameter 40 mm the amplitude needed for cavitation onset is approximately 0.3 µm, with a 20 mm diameter, the amplitude changes to approximately 0.6 µm (Figure 4). Therefore the power radiated per unit of area approximately doubles.

In order to increase the reproducibility when the sonication is close to the cavitation threshold the use of the power irradiated is not suitable for many common probes and frequencies. The dimensionless number $G_A$ represents a better option in these cases.

### 6. Conclusions.

A dimensionless number that predicts cavitation in accelerated systems has been used to predict ultrasonic cavitation.

The dimensionless number is

$$G_A = \frac{p - p_v}{\rho a L} \qquad (7)$$

where $G_A$ is the dimensionless number, $p$ is the pressure of the liquid at the depth of the probe tip, $p_v$ is the vapour pressure of the liquid, $\rho$ is the density of the liquid, $L$ is the shortest between the radius of the probe and half the wavelength, and $a$ is the acceleration of the tip.

For values of $G_A \ll 1$ cavitation is expected; and for values of $G_A \gg 1$ cavitation is not expected.

The power irradiated per unit of area is not a good parameter to predict cavitation onset in many common applications, i.e. low ultrasonic frequency and medium diameters. $G_A$ represents a better parameter to predict cavitation and to increase reproducibility when the conditions are close to the cavitation threshold.

# References

1. Franc J, Michel J. Introduction the Main Features of Cavitating Flows. Fundamentals of Cavitation. 2005:1-14.

2. Tuulmets A, Piiskop S, Järv J, Salmar S. Sonication effects on non-radical reactions. A sonochemistry beyond the cavitation? Ultrason Sonochem. 2014;21(3):997-1001.

3. Yu D, Liu B, Wang B. The effect of ultrasonic waves on the nucleation of pure water and degassed water. Ultrason Sonochem. 2012;19(3):459-63.

4. Pitt WG, Ross SA. Ultrasound increases the rate of bacterial cell growth. Biotechnol Prog. 2003;19(3):1038-44.

5. Yachmenev V, Condon B, Klasson T, Lambert A. Acceleration of the enzymatic hydrolysis of corn stover and sugar cane bagasse celluloses by low intensity uniform ultrasound. Journal of Biobased Materials and Bioenergy. 2009;3(1):25-31.

6. Chisti Y. Sonobioreactors: using ultrasound for enhanced microbial productivity. Trends Biotechnol. 2003;21(2):89-93.

7. Holland CK, Roy R, Apfel R, Crum L. In vitro detection of cavitation induced by a diagnostic ultrasound system. IEEE Trans Ultrason Ferroelectr Freq Control. 1992;39(1):95-101.

8. Cravotto G, Cintas P. Power ultrasound in organic synthesis: moving cavitational chemistry from academia to innovative and large-scale applications. Chem Soc Rev. 2006;35(2):180-96.

9. Miller MW, Miller DL, Brayman AA. A review of in vitro bioeffects of inertial ultrasonic cavitation from a mechanistic perspective. Ultrasound Med Biol. 1996;22(9):1131-54.

10. Mason TJ, Peters D. 1.4 Parameters that affect cavitation. In: Practical sonochemistry: Power ultrasound uses and applications. Woodhead Publishing; 2002. p. 8.

11. Hodnett M, Choi MJ, Zeqiri B. Towards a reference ultrasonic cavitation vessel. Part 1: Preliminary investigation of the acoustic field distribution in a 25kHz cylindrical cell. Ultrason Sonochem. 2007;14(1):29-40.

12. Wang L, Memoli G, Hodnett M, Butterworth I, Sarno D, Zeqiri B. Towards a reference cavitating vessel Part III? design and acoustic pressure characterization of a multi-frequency sonoreactor. Metrologia. 2015;52(4):575.

13. Garcia-Atance Fatjo G. New Dimensionless Number To Predict Cavitation In Accelerated Fluid. International Journal of Computational Methods and Experimental Measurements. 2016;4(4):484-92.